\documentclass[12pt]{article}
\usepackage{siunitx}
\usepackage{graphicx}
\usepackage[utf8]{inputenc}
\usepackage{graphicx,xcolor,setspace,geometry}
\usepackage{amsmath,setspace}
\usepackage{mathtools}
\usepackage{multirow}
\usepackage{booktabs}
\usepackage{threeparttable}
\usepackage{adjustbox}
\usepackage{nicefrac}
\usepackage{caption}
\title{Continuous--time multi--state \\ capture--recapture models}

\author{Sina Mews$^{1,}\footnote{Corresponding author; email: \texttt{sina.mews@uni-bielefeld.de}.}$, Roland Langrock$^{1}$, Ruth King$^{2}$, and Nicola Quick$^{3,4}$ \\
 \\
$^{1}$Bielefeld University, Germany \\
$^{2}$University of Edinburgh, UK \\
$^{3}$University of St Andrews, UK \\
$^{4}$Duke University Marine Lab, U.S.A.}

\date{}

\begin{document}

\begin{spacing}{1.25}
    \maketitle
\end{spacing}

\vspace{-10mm}

\begin{spacing}{1.5}

\begin{abstract}
Multi-state capture-recapture data comprise individual-specific sighting histories together with information on individuals' states related, for example, to breeding status, infection level, or geographical location. 
Such data are often analysed using the Arnason-Schwarz model, where transitions between states are modelled using a discrete-time Markov chain, making the model most easily applicable to regular time series. 
When time intervals between capture occasions are not of equal length, more complex time-dependent constructions may be required, increasing the number of parameters to estimate, decreasing interpretability, and potentially leading to reduced precision.
Here we develop a novel continuous-time multi-state model that can be regarded as an analogue of the Arnason-Schwarz model for irregularly sampled data. 
Statistical inference is carried out by regarding the capture-recapture data as realisations from a continuous-time hidden Markov model, which allows the associated efficient algorithms to be used for maximum likelihood estimation and state decoding.
To illustrate the feasibility of the modelling framework, we use a long-term survey of bottlenose dolphins where capture occasion are not regularly spaced through time.
Here we are particularly interested in seasonal effects on the movement rates of the dolphins along the Scottish east coast. 
The results reveal seasonal movement patterns between two core areas of their range, providing information that will inform conservation management. 
\end{abstract}

\noindent \textbf{Keywords:}
Arnason-Schwarz model, continuous-time Markov chain, hidden Markov model, maximum likelihood.

\section{Introduction}
\label{s_intro}

\subsection{Multi-State Capture-Recapture: Discrete vs.\ Continuous Time}

Capture-recapture studies use repeated surveys of a population of interest to infer properties of the underlying ecological system.
On several survey occasions, all individuals observed are (re-)captured, identified based on individual marks, and subsequently released back into the population.
In the simplest case, the resulting individual-specific capture histories provide information on the presence or absence of the individual at each occasion.
In some studies, additional information on the discrete state of the individual at the time of the capture event is also recorded. 
This information may correspond to geographical locations (e.g.\ migration between different areas; Schwarz, Schweigert, and Arnason, 1993; Brownie et al., 1993; King and Brooks, 2004; Worthington et al., 2019), reproductive status (e.g.\ breeding vs.\ not breeding; Pradel and Lebreton, 1999; McCrea et al., 2010), or health status (e.g.\ infected vs.\ not infected; Faustino et al., 2004; Conn and Cooch, 2009).
In those instances, \textit{multi-state} capture-recapture models are often used to explore individuals' transitions between these different states.
The Arnason-Schwarz (AS) model --- a generalisation of the Cormack-Jolly-Seber (CJS) model from the single- to the multi-state case --- is the classic modelling framework for analysing corresponding multi-state capture-recapture data (Arnason, 1973; Schwarz et al., 1993; Brownie et al., 1993). 
Within the standard AS model, the state process is assumed to follow a discrete-time first-order Markov chain, although King and Langrock (2016) extend this to the semi-Markov case for live state transitions.
    
Due to the discrete-time formulation of the AS model, the model in its basic form is best suited to capture occasions that follow a regular sampling protocol (e.g.\ with monthly or yearly capture occasions but allowing for missed capture occasions at some of these times). 
Specifically, inference within discrete-time models, for example with respect to how the states evolve over time, is to be seen relative to the fixed interval length between capture occasions. 
For example, when inferring survival rates from encounter histories with yearly capture occasions, the corresponding estimates relate to the probability of individuals surviving one year.  
Therefore, discrete-time model formulations are typically inadequate when time intervals between capture occasions are irregular (i.e.\ not equidistant in time).
In some cases, the corresponding continuous process can be transferred into a discrete process by temporal aggregation. 
However, this would introduce subjectivity regarding the choice of the discrete-time modelling resolution and a loss of information (Borchers et al., 2014). 
Moreover, applying discrete-time models to capture-recapture data with irregular sampling occasions has been shown to lead to biased estimates (Yip and Wang, 2002; Barbour, Ponciano, and Lorenzen, 2013).
Continuous-time model formulations are conceptually superior for addressing irregularly spaced encounter histories, but are mathematically more challenging than their discrete-time counterparts.
    
For closed populations, i.e.\ populations unaltered by births, deaths, and migration, continuous-time models have been developed and extended for decades (see, e.g., Becker, 1984; Chao and Lee, 1993; Yip, Huggins, and Lin, 1996; Hwang and Chao, 2002; Schofield, Barker, and Gelling, 2018). 
There has been recent renewed interest in such continuous-time models due to technological advances in data collection within capture-recapture studies. 
In particular, arrays of motion-sensor cameras or acoustic sensors that record individuals in continuous time, rather than at pre-determined (static) capture occasions, have led to the development of continuous-time capture-recapture models for spatially explicit capture-recapture studies (Borchers et al., 2014; Distiller and Borchers, 2015; Kolev et al., 2019). 
In contrast, for open populations, only very few contributions in the literature consider continuous-time modelling approaches.
Recently, Fouchet et al.\ (2016) developed a continuous-time model for the estimation of survival rates based on capture-recapture data. 
Their model, which comprises only two states, namely the animal being alive or dead, is the continuous-time analogue of the CJS model.
To our knowledge, Choquet et al.\ (2017) and Choquet (2018) were the first to use a more general multi-state continuous-time modelling framework. 
In their analyses, opportunistic data, collected via citizen science without any sampling design, are modelled using a Markov-modulated Poisson process, which assumes that observation occasions are drawn from a Poisson process. 
The approach is illustrated for opportunistic capture-recapture data on the Alpine ibex (\textit{Capra ibex}), where individuals could be observed at any time throughout the study period.
    
In this contribution, we propose a continuous-time AS (i.e.\ multi-state) model that naturally addresses any irregular sampling.
The key difference between our contribution and those of Choquet et al.\ (2017) and Choquet (2018) lies in the nature of the capture occasions: while in those previous contributions the times of the capture occasions 
were regarded as realisations of a Poisson process, we treat them as fixed (but irregularly spaced in time). 
We regard the capture-recapture setting as a special case of a (partially) hidden Markov model (HMM) in continuous time, where an individual's capture history corresponds to the observed state-dependent process and the individual's (true) state corresponds to the (partially observed) state process (cf.\ Pradel, 2005; Gimenez et al., 2012; King and McCrea, 2014; King and Langrock, 2016).
Continuous-time HMMs have been used in other applications before, most notably in medical statistics (see, e.g., Jackson et al., 2003; Conn and Cooch, 2009; Alaa and van der Schaar, 2018; Amoros et al., 2019; Williams et al., 2019).
The continuous-time HMM framework allows us to exploit the efficient HMM-based forward algorithm for parameter estimation, as well as the wider standard HMM machinery, for example for decoding the underlying states.
We note that not only the transition rates between states are of interest, but also potential covariate effects on these transitions.
We demonstrate how transition rates can be modelled as a function of covariates, which may be individual-specific, time-varying, or both. 
For situations with time-varying covariates, the likelihood is analytically intractable, but can be approximated using piecewise constant functions, as suggested for example by Lebovic (2011) and Langrock, Borchers, and Skaug (2013).

\subsection{Motivating Example: Annual Movement Patterns of Bottlenose Dolphins}

The east coast of Scotland is home to a small, resident bottlenose dolphin (\textit{Tursiops truncatus}) population, that ranges large distances up and down the coast. 
Human activity, including oil and gas exploration and large-scale offshore wind farm development, overlaps the core range of this dolphin populations creating the potential for disturbance. 
To better identify potential effects, understanding how individuals move along the coast will aid in decision making and minimise any potential negative impacts on the dolphin population.
    
Survey effort to document individual dolphins has taken place since 1990, providing one of the longest running studies on bottlenose dolphins anywhere in the world.  
Effort has been focused in two main areas: the Moray Firth since 1990 (including the Special Area of Conservation; this region is labelled as SAC in the following) and Tayside \& Fife since 1997 (T\&F). 
In both areas, photo-identification data is collected from small boats during periods of favourable weather. 
Photo-identification is widely used in cetacean research (W\"ursig and W\"ursig, 1977) to discriminate between individuals using long-lasting natural marks.
Using photographs, individual encounter histories are constructed, providing information on individual movement between SAC and T\&F.
In total, sighting histories of 835 bottlenose dolphins from a total of 1,110 boat trips (SAC: 998 trips; T\&F: 201 trips) were considered.
Individuals were excluded from analysis if they were sighted less than 6 times in total and if their sex was unknown.
Our analysis is hence based on 207 individuals (95 males and 112 females) with a median number of 45 sightings (min: 6 sightings; max: 242 sightings).

\begin{figure}[!htb]
    \centering
    \includegraphics[width=102mm]{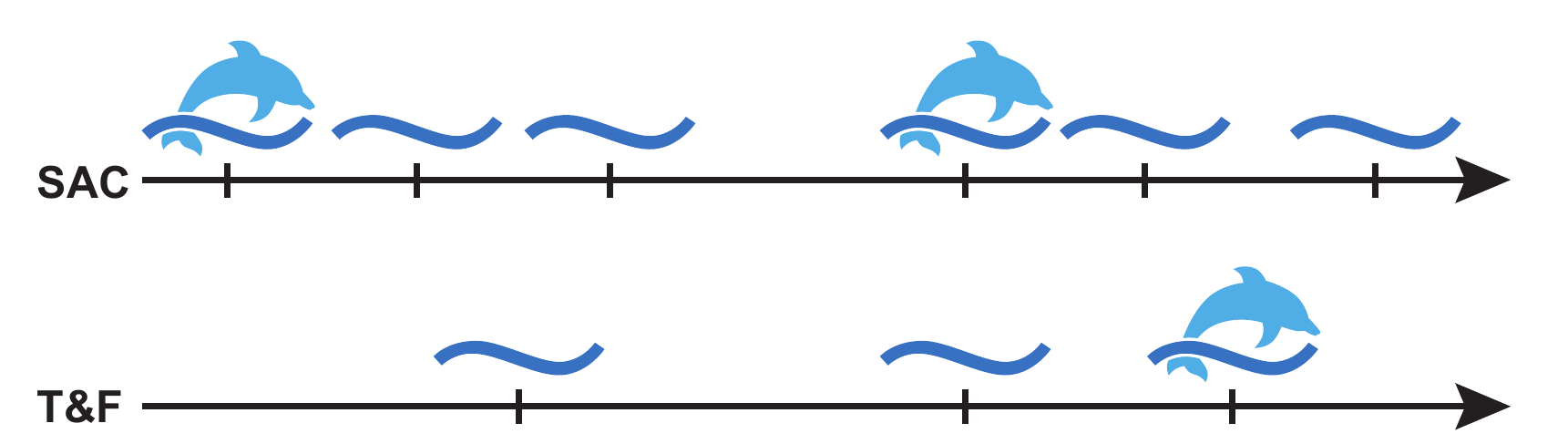}
    \caption{Graphical illustration of a possible encounter history for one dolphin. The waves indicate time points at which capture occasions occurred, while the dolphin is only observed at some of those occasions (two times in SAC and one time in T\&F).}
    \label{fig_capHist}
\end{figure}

The key challenge of these data is the lack of a regular sampling interval, with capture occasions irregularly spaced in time due to constraints posed by weather conditions.
Most capture occasions occur between May and the beginning of October, whilst almost none occur during the winter months.
A graphical illustration of a corresponding encounter history is given in Figure~\ref{fig_capHist}. 
In contrast to commonly conducted capture-recapture studies with equidistant times between capture occasions, the interval length between occasions in our study is not fixed but varies temporally from a few days to several weeks.   
As a consequence, each capture occasion needs to be supplemented with information on the time it took place.
The resulting encounter history associated with Figure~\ref{fig_capHist} would thus look as follows: \vspace{-3mm} \\
    
\begin{tabular}{lcccccccc}
     encounter history: & 1 & 0 & 0 & 0 & 1 & 0 & 2 & 0 \\
     days since first capture: & 0 & 5 & 8 & 10 & 23 & 28 & 29 & 33
\end{tabular} \\ 

\noindent At each capture occasion, the individual can be observed in either SAC or T\&F (denoted by 1 or 2, respectively); or, in the most common case, the dolphin is not observed (denoted by 0).

In Section \ref{s_model}, we first provide the general formulation of a multi-state capture-recapture model in continuous time. 
Subsequently, in Section \ref{s_apply}, we present the results of fitting such a model to the bottlenose dolphin data, and investigate properties of the proposed inferential approach in simulation experiments in Section \ref{s_sim}.
A final discussion of our modelling approach is given in Section \ref{s_discuss}.

\section{Continuous--Time Multi--State Capture--Recapture Model}
\label{s_model}

\subsection{Basic Model Formulation}

We consider encounter histories of individuals which, when captured alive, are encountered in one of finitely many states (in our case study corresponding to the two different sites). 
Let $n$ denote the total number of individuals observed, $T$ the total number of capture occasions and $\mathcal{M}=\{1,\ldots,M\}$ the set of possible states while alive. 
Then for each individual $i=1,\ldots,n$, at capture times $t=t_0,t_1,\ldots,t_T$, where $0=t_0<t_1<\ldots<t_T$, the observed event is given by
$$ x_{i,t}=
	\begin{cases}
	0 & \text{if individual $i$ is not observed at time $t$;}\\
    m & \text{if individual $i$ is observed in state $m$ at time $t$.}
	\end{cases}$$
The individual capture histories, denoted by ($x_{i,t_0},\ldots,x_{i,t_T}$), suffer from imperfect detection, such that we formulate our model as a hidden Markov model (HMM), i.e.\ a state-space model with discrete state space (see, e.g., Buckland et al., 2004; Gimenez et al., 2007; Royle, 2008; Schofield and Barker, 2008; King, 2014). 
State-space models consist of two parallel processes, a (partially observed or unobserved) state process and an observation process, conditional on the state of the individual. 
Using this model formulation allows us to distinguish the observation process (i.e.\ the recapture process, which corresponds to the individual encounter histories) from the underlying state process (i.e.\ the demographic process of interest, in our setting including the distinct alive states of interest). 
For capture occasions at which the individual is encountered, we have information on the individual's state, whereas the state is unknown when the individual is not encountered. 
The state process is thus partially observed. 
We define the state process ($s_{i,t_0},\ldots,s_{i,t_T}$) such that
$$ s_{i,t}=
	\begin{cases}
    m & \text{if individual $i$ is alive and in state $m$ at time $t$;}\\
    M+1 & \text{if individual $i$ is dead at time $t$}.
	\end{cases}$$
The states $1,\ldots,M$ correspond to a categorical variable of interest, in our motivating example the location of an individual (with $M=2$).
For convenience we will drop the subscript $i$ from now on, but will continue to refer to the individual. 

The observation process is characterised by the state-specific recapture (or detection) probabilities $p_m$, which are generally defined as $p_m = \text{Pr}(x_t=m | s_t=m)$ for $t=t_1,\ldots,t_T$.
The recapture probabilities may depend, \textit{inter alia}, on (state-specific) survey effort, which in our motivating example corresponds to whether a state, i.e.\ the area of interest, was searched at a capture occasion or not.
Overall, we have the following probabilities associated with the different events that can occur at a given capture occasion:
$$ \text{Pr}(x_t|s_t)=
	\begin{cases}
    p_m & \text{for} \quad x_t=m, s_t=m; \\
    1-p_m & \text{for} \quad x_t=0, s_t=m; \\
    1 & \text{for} \quad x_t=0, s_t=M+1; \\
    0 & \text{otherwise};
	\end{cases} $$
for $t=t_1,\ldots,t_T$. 
Note that the probabilities of observing the different events depend on the individual's state $s_t$ at survey occasion $t$, which is only partially known. 
Furthermore, we assume that information on the states is collected without error, i.e.\ $\text{Pr}(x_t=m|s_t\neq m)=0$. 
However, this can be extended to allow for additional observation error using ideas similar to, for example, Pradel (2005) and King and McCrea (2014).

We now turn to the mechanisms of the state process. 
Due to the temporal irregularity of the survey occasions, there is no natural discrete-time model formulation for the state process. 
For such a time series structure the model parameters would have to be interpreted with respect to a fixed interval length (i.e.\ sampling unit, such as one month) between occasions. 
Instead, we use a continuous-time Markov chain, $\{S_t\}_{t \geq 0}$, $S_t \in \{1,\ldots,M,M+1 \}$, to model how the states evolve over time. 
According to the Markov property, conditional on the individual's state at time $u$, its state at any time $v$ with $v> u$ is then independent of the trajectory of states prior to $u$.
The transitioning between the different states is governed by an underlying transition intensity matrix,
$$\mathbf{Q}=\begin{pmatrix}
q_{1,1} & \ldots & q_{1,M} & q_{1,M+1}\\
\vdots & \ddots & \vdots & \vdots\\
q_{M,1} & \ldots & q_{M,M} & q_{M,M+1}\\
0 & \ldots & 0 & 0
\end{pmatrix},$$
where the state transition intensities are defined as 
$$q_{j,k}=\lim\limits_{\triangle t \to 0}\frac{\text{Pr}(s_{t+\triangle t}=k | s_t=j)}{\triangle t}.$$ 
These can be interpreted as the rate at which transitions from state $j$ to state $k$ occur. 
Due to the constraints that $q_{j,k} \geq 0$ for $j \neq k$ and $\sum_{k=1}^{M+1} q_{j,k} = 0$, the diagonal entries are obtained as $ q_{j,j} = -\sum_{k \neq j} q_{j,k} $. 
The last row of the intensity matrix $\mathbf{Q}$ consists of zeros only because we assume the last state, corresponding to an individual's death, to be an absorbing state from which no transition back to the other states is possible. 
The sojourn time in each state $j=1,\ldots,M$ is exponentially distributed with parameter $-q_{j,j}$, leading to a mean sojourn time of $-1/q_{j,j}$. 
We initially consider the transition intensities to be constant over time, which we will relax in Section \ref{sub_covariates} when including time-varying covariates.
Given a time-homogeneous intensity matrix $\mathbf{Q}$, the transition probability matrix (t.p.m.) $\boldsymbol{\Gamma}$ for a time interval between two consecutive capture occasions, $[t_{u-1},t_u]$, $u=1,\ldots,T$, is then obtained as a matrix exponential (Cox and Miller, 1965),
\begin{equation}
    \boldsymbol{\Gamma}(t_{u-1},t_u) = \text{exp}\bigl\{\mathbf{Q}\cdot(t_u-t_{u-1})\bigr\} = \sum_{d=0}^\infty \mathbf{Q}^d(t_u-t_{u-1})^d / d!.
\label{tpm}
\end{equation}
The entries $\gamma_{j,k}(t_{u-1},t_u)$ in this t.p.m.\ indicate the probability to move from state $j$ at capture occasion $t_{u-1}$ to state $k$ at the next capture occasion $t_u$.

\subsection{Likelihood Evaluation and Maximisation}

In the following we let $g \in \{0,\ldots,T-1\}$ denote the capture occasion at which a given individual was first sighted (hence at time $t_g$). 
Let $W=\{ u \in \{ g, g+1,\ldots,T \} \,|\, x_{t_u} \in \mathcal{M}\}$ denote the set of indices of the capture occasions at which the individual was observed, such that its state at the corresponding times is \textit{known}. 
The complement $W^c = \{g,g+1,\ldots,T\}\backslash W$ is the set of capture occasions at which the individual's state is \textit{unknown}.
Conditional on the initial capture at time $t_g$, and in particular --- since we assume the states to be recorded whenever an individual is captured --- conditional on the state $s_{t_g}$, the likelihood of the individual's capture history is then given as
\begin{equation}
    \mathcal{L} 
    = \sum_{\tau \in W^c} \sum_{s_{t_\tau} \in \{1,\ldots,M,M+1\}}  \prod_{u=g+1}^T \gamma_{s_{t_{u-1}},s_{t_u}}(t_{u-1},t_u) \text{Pr}(x_{t_u} | s_{t_u}).
\label{llk_sum}
\end{equation}
In the likelihood calculation, we sum over all possible state sequences which are compatible with the observed capture history. 
Note that here we have made the additional assumption that conditional on the state process, the observations (i.e.\ the recaptures) are independent of each other --- in other words, the probability of an individual being recaptured is completely determined by its state, and hence not additionally affected by potential previous recaptures.

Equation (\ref{llk_sum}) corresponds to a brute force calculation of the likelihood of an individual's capture history, where at each time point with unknown state we sum over all possible states.
In the worst case, the computational cost of evaluating the expression given in (\ref{llk_sum}) is of order $\mathcal{O}\bigl(T(M+1)^T\bigr)$ (if the individual considered was not encountered anymore after the initial capture).
Since we formulate the capture-recapture model within an HMM framework, we can however use the corresponding recursive techniques to more efficiently evaluate the likelihood.
Following Jackson et al.\ (2003), we use the HMM forward algorithm to calculate the likelihood, yielding the matrix product
\begin{equation}
    \mathcal{L} = 
    \Bigl[ \prod_{u=g+1}^T \underbrace{\text{exp} \bigl\{\mathbf{Q}\cdot(t_u-t_{u-1}) \bigr\}}_{\mathclap {= \boldsymbol{\Gamma}(t_{u-1},t_u) }} \mathbf{P}(x_{t_u}) \Bigr] \boldsymbol{1},
\label{llk}
\end{equation}
where $\mathbf{P}(x_t) = \text{diag}\bigl\{ \Pr(x_t | s_t=1), \ldots , \Pr(x_t | s_t=M+1) \bigr\}$ is a diagonal matrix, and where $\boldsymbol{1}$ is a column vector of length $M+1$ of ones.
In the worst case, the computational cost of evaluating (\ref{llk}) is of order $\mathcal{O}\bigl(T (M+1)^2\bigr)$ only.

Assuming independence of the capture histories over individuals, conditional on the model parameters, the likelihood over multiple capture histories is simply calculated as the product of the individual likelihoods $\mathcal{L}$ given in (\ref{llk}). 
The model parameters, namely the state transition intensities as well as the detection probabilities, are then estimated by numerically maximising the joint likelihood, subject to the usual technical problems including local maxima and parameter constraints (for a detailed account of how to address these issues, see, for example, Zucchini, MacDonald, and Langrock, 2016).

\subsection{Incorporating Time--Varying Covariates}
\label{sub_covariates}

In general, and in particular in our motivating example --- where we are interested in seasonal variations in movements between the two sites of interest --- the state transition intensities may depend on some time-varying covariate $h(t)$, e.g.\ such that $q_{j,k}(t) = \text{exp} \bigl\{ \alpha_{jk0} + \alpha_{jk1} h(t) \bigr\}$, for $j \ne k$.
However, incorporating such covariates into the continuous-time state process is rather challenging: Equation (\ref{tpm}) no longer holds for non-homogeneous transition intensities, and as a consequence the likelihood function becomes intractable. 
An important exception is the case where the covariate of interest and hence also the intensities are piecewise constant over time (see, e.g., Faddy, 1976; Kay, 1986; Lebovic, 2011; Langrock et al., 2013).
We thus partition the time interval during which observations are made, $[0,t_T]$, into $R$ intervals, $\tau_1,\ldots,\tau_R$, with $\tau_r = [b_{r-1},b_r)$ and $b_0=0$, $b_R=t_T$.
For simplicity, we consider intervals of constant length $l=b_1-b_0=\ldots=b_R-b_{R-1}$, on which the (potentially continuously varying) transition intensities are approximated by a constant function. 
This approximation leads to a simple closed-form expression of the likelihood, without the need to evaluate integrals. 
Specifically, for piecewise constant transition intensities, we obtain the t.p.m. $\boldsymbol{\Gamma}(t_{u-1},t_u)$ within (\ref{llk}) recursively as a product of t.p.m.s associated with intervals over which the intensities are constant (a consequence of the Chapman-Kolmogorov equations).
In general, for each individual the likelihood contribution is
\begin{equation}
\label{llk_cov}
    \mathcal{L} = \Bigl[ \prod_{u=g+1}^T \boldsymbol{\Gamma}(t_{u-1},t_u) \mathbf{P}(x_{t_u}) \Bigr] \boldsymbol{1},
\end{equation}
        where
    $$ \boldsymbol{\Gamma}(t_{u-1},t_u) = \begin{cases}
    \text{exp} \bigl\{\mathbf{Q}_r(t_u-t_{u-1})\bigr\} & \text{if $t_{u-1}, t_u \in \tau_r $,} \vspace{2mm}\\
    \text{exp} \bigl\{\mathbf{Q}_r(b_r-t_{u-1})\bigr\} \left[ \prod_{v=r+1}^{s-1} \text{exp} \bigl\{\mathbf{Q}_v(b_v-b_{v-1})\bigr\} \right] \\[-0.8em] \qquad \times \text{exp} \bigl\{\mathbf{Q}_s(t_u-b_{s-1})\bigr\} & \text{if $t_{u-1} \in \tau_r, t_u \in \tau_s$,}
	\end{cases} $$
with $\mathbf{Q}_r$, $r=1,\ldots,R$, denoting the constant (approximating) intensity matrix over the interval $\tau_r$, and using the convention that the empty product is equal to 1 (if $s=r+1$).
The approximation of the time-varying intensities by step functions thus allows us to estimate the parameters by numerically maximising a likelihood similar to (\ref{llk}), which is an approximation of the likelihood of the actual model of interest. 
Crucially, the approximation can be made arbitrarily accurate by decreasing the width of the intervals. The effect of the approximation will be investigated in Section \ref{s_sim}.

\section{Case Study: Movement Patterns of Bottlenose Dolphins}
\label{s_apply}

\subsection{Model Formulation}

We return to our motivating data, and specify $M=2$ states while alive, corresponding to the dolphin's current location (i.e.\ the area SAC or T\&F). 
As capture occasions can occur in either SAC or T\&F, or in fact both at the same time, we need to consider which area is searched at a given occasion at time $t$. 
This is indicated by the dichotomous variable $a^{(m)}_t$ (corresponding to the state-specific survey effort), with $a_t^{(m)}=1$ if the area associated with state $m$ was surveyed at time $t$, and $a_t^{(m)}=0$ otherwise.
The conditional distribution of the state-dependent process can thus be summarised as follows when considering the state-specific survey effort:
$$ \text{Pr}(x_t|s_t)=
	\begin{cases}
    p_m a^{(m)}_t & \text{for} \quad x_t=m, s_t=m \\
    1 - p_m a^{(m)}_t & \text{for} \quad x_t=0, s_t=m \\
    1 & \text{for} \quad x_t=0, s_t=M+1; \\
    0 & \text{otherwise};
	\end{cases} $$
for $t=t_1,\ldots,t_T$. 
In particular, if the area associated with state $m$ is not surveyed at capture occasion $t$, the probability of recapture for that area is zero (i.e.\ $\text{Pr}(x_t = m|s_t = m) = 0$ if $a^{(m)}_t=0$).

Within the modelling of the annual movements of bottlenose dolphins between SAC and T\&F, we also want to account for potential sex differences. 
Our covariates of interest thus are time of year, denoted by $y(t)$, and sex, denoted by the binary variable $z$ (with $z=0$ for females and $z=1$ for males).
Moreover, we include interactions between time of year $y(t)$ and sex $z$ in our model, hence allowing for different migration patterns of females and males over time.
Given the binary nature of $z$, this is equivalent to estimating separate parameters for females and males for the effect of time of year $y(t)$.
Thus, for both females and males, we model the (off-diagonal) transition intensities as a function of the time of year $y(t)$ using trigonometric functions to accommodate the periodicity:
\begin{equation}
\label{TransRates}
    q^{(z)}_{j,k}(t) = \text{exp} \left\{ \alpha^{(z)}_{jk0} + \alpha^{(z)}_{jk1} \text{sin}\Bigl( \frac{2 \pi y(t)}{365} \Bigr) + \alpha^{(z)}_{jk2} \text{cos}\Bigl( \frac{2 \pi y(t)}{365} \Bigr) \right\},
\end{equation}
with parameter vector $\boldsymbol{\alpha}^{(z)}_{jk}=(\alpha^{(z)}_{jk0}, \alpha^{(z)}_{jk1}, \alpha^{(z)}_{jk2})$, to be estimated for $j,k=1,2$, $j\neq k$.

As described in Section \ref{sub_covariates}, the corresponding likelihood is intractable.
Therefore, we divide the study period of 10,486 days into intervals of constant length $l=30$ days, which provided a good balance between approximation accuracy and computational cost.
Within the resulting intervals $\tau_1,\ldots,\tau_{350}$, the continuously varying $y(t)$ in Equation (\ref{TransRates}) is then approximated by the midpoint $c_r$ of the corresponding interval $\tau_r$:
\begin{equation}
\label{piecewiseTransRates}
    q^{(z)}_{j,k}(t) = \text{exp} \left\{ \beta^{(z)}_{jk0} + \beta^{(z)}_{jk1} \text{sin}\Bigl( \frac{2 \pi c_r}{365} \Bigr) + \beta^{(z)}_{jk2} \text{cos}\Bigl( \frac{2 \pi c_r}{365} \Bigr) \right\}, \text{ for }  t \in \tau_r,
\end{equation}
with parameter vector $\boldsymbol{\beta}^{(z)}_{jk}=(\beta^{(z)}_{jk0}, \beta^{(z)}_{jk1}, \beta^{(z)}_{jk2})$, to be estimated for $j,k=1,2$, $j\neq k$.

In contrast to time-varying covariates, it is straightforward to model the effect of a discrete, individual-specific covariate like sex on the transition intensities.
In our case, we allow for differences in the (apparent) mortality rates between males and females, i.e.\ the transition intensities into state 3 (corresponding to the individual being dead), while we assume the death rates to be the same for both sites and to be constant over time.
Therefore, the mortality rates are modelled as
\begin{equation*}
\label{deathRates}
    q_{j,3} = \beta_{30} + \beta_{31} z, \quad \text{for} \quad j=1,2.
\end{equation*}
The parameter of interest here is $\beta_{31}$, which indicates the difference in (apparent) mortality, and hence survival, between the sexes.

\subsection{Results}
\label{ss_results}

\begin{figure}[!hb]
    \centering
    \includegraphics[width=110.8mm]{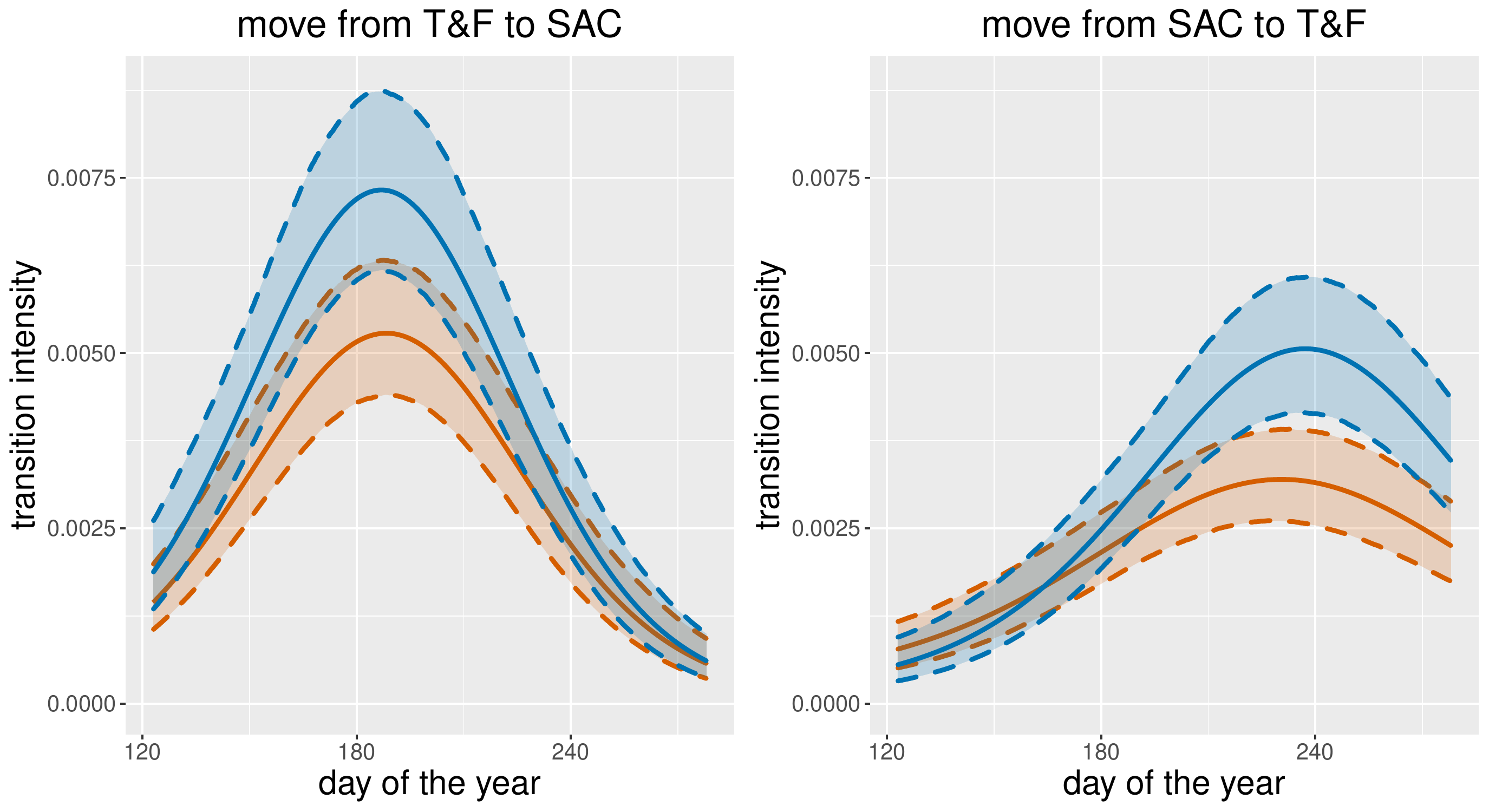}
    \caption{Estimated transition intensities (mean and 95\% CI) as a function of the covariate \textit{day of the year} plotted for the period from May to October. Left plot shows the intensities to move from T\&F to SAC and right plot vice versa. Blue is for male bottlenose dolphins and orange for female ones. CIs for the intensities were obtained based on Monte Carlo simulation from the estimators’ approximate distribution as implied by maximum likelihood theory. Coefficients underlying this figure are provided in Table~\ref{tab_resSI} in Appendix~A.}
    \label{fig_res}
\end{figure}

\begin{figure}[!hb]
    \centering
    \includegraphics[width=115mm]{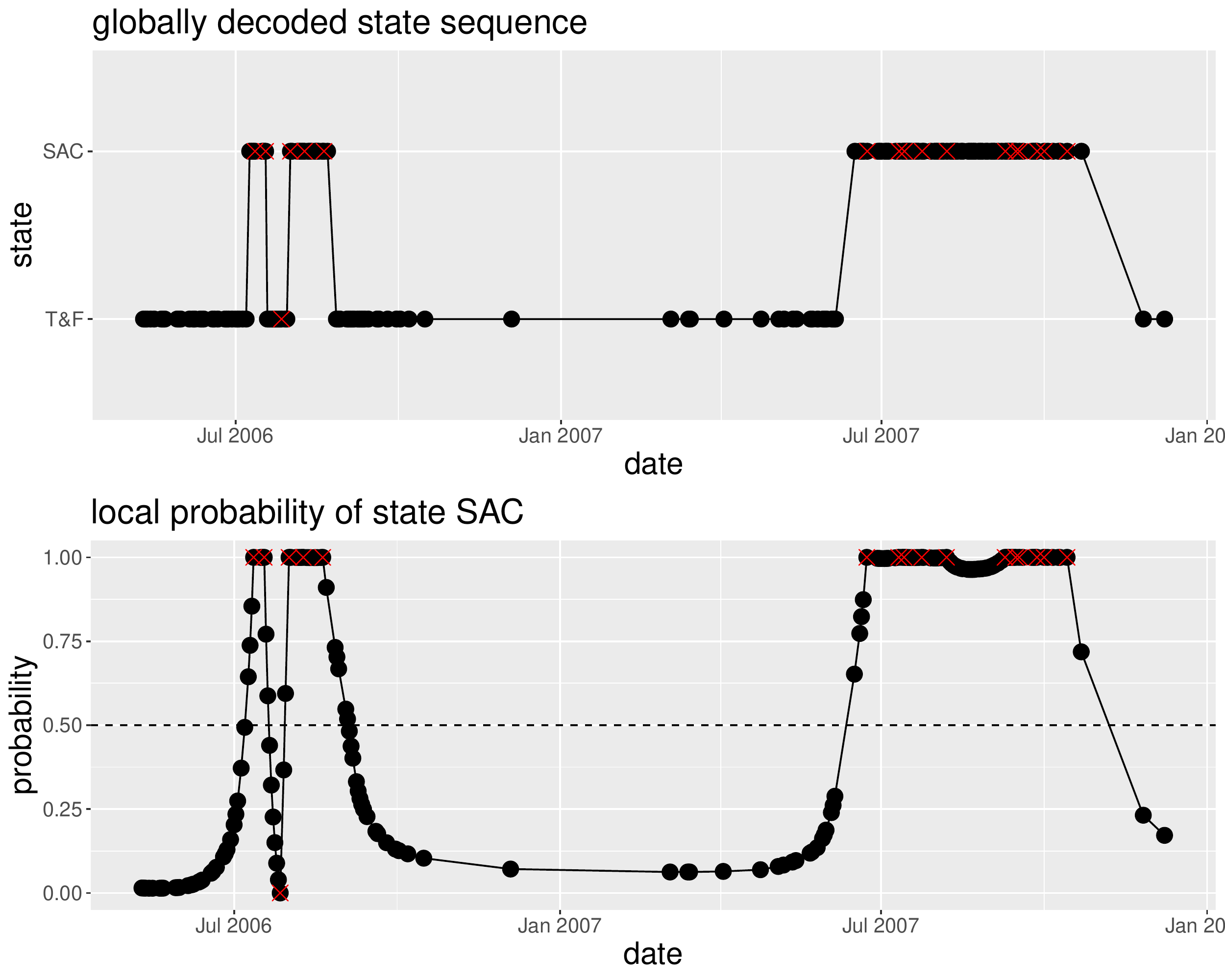}
    \caption{Example of a decoded state sequence for one male bottlenose dolphin. The upper plot shows the globally decoded states, while the lower plot shows the probabilities of state SAC for each capture occasion. The red crosses indicate recapture of the dolphin.}
    \label{fig_decStates}
\end{figure}

The estimation results reveal clear seasonal patterns with higher intensities of movement from T\&F to SAC in summer, whereas intensities to move from SAC to T\&F are highest in autumn (see Figure~\ref{fig_res}).
Moreover, male bottlenose dolphins have generally higher transition intensities than females, leading to mean sojourn times for the highest estimated intensities of 305 days in SAC (186 in T\&F) for females and 194 days in SAC (135 in T\&F) for males, respectively. 
These figures suggest that movement between the two sites is infrequent, which is also reflected in the (individual-specific) most likely state sequences, given the encounter histories and the fitted model. 
These sequences provide individual information on the (apparent) survival status as well as the spatial position throughout the observation period. 
An example of such a state sequence, decoded using the Viterbi algorithm and complemented with the local state probabilities as obtained using the forward-backward algorithm, is shown in Figure~\ref{fig_decStates} for one of the male dolphins.
This kind of probabilistic inference on the only partially observed state process is an example of how the HMM framework can be utilised to supplement information from encounter histories, in this case by inferring the movement between sites even at times when the individual was only rarely observed (the recapture probabilities are estimated around 20\% for both areas).
Furthermore, the decoded states can provide information on when a dolphin may have died (or left the study area).
In general, our model indicates that males have higher mortality rates than females, with expected (apparent) survival times of about 33 years for females and just under 31 years for males, although this difference was not found to be significant (cf.\ Table~\ref{tab_resSI} in Appendix~A).

\section{Simulation Experiments}
\label{s_sim}

Simulations were conducted to explore the effect of approximating continuously varying transition intensities by piecewise constant intensities, in particular with regard to the resulting estimator properties.
It is intuitively clear that the smaller the intervals used for the approximation, the more reliable the estimation, but the longer the computation time --- a classical trade-off situation. 
We thus want to develop some intuition regarding which interval lengths used in the approximation serve as a good compromise between estimation accuracy and computational cost.

The simulation setting is based on our motivating data.
We consider simulated capture histories of $n=200$ individuals with $T=620$ capture occasions over ten years, from $t_0=0$ to $t_{620}=3,646$ days. 
We consider $M=2$ alive states and choose the state-specific detection probabilities as $p_1=0.4$ and $p_2=0.2$, respectively. 
The intervals between survey occasions were drawn from Poisson distributions with means $\lambda_1=10$ for state 1 and $\lambda_2=14$ for state 2, respectively. 
As in our motivating example, we model the (off-diagonal) transition intensities as a function of the time of year $y(t)$ using Equation (\ref{TransRates}).
Since the focus is on the likelihood approximation in the presence of time-varying covariates, we do not include sex as a covariate in our simulations.
We thus have only one set $\boldsymbol{\alpha}_{jk}$ for $j,k=1,2$ and $j \neq k$, instead of two distinct sets (one for males and one for females).
The parameters $\boldsymbol{\alpha}_{jk}$, for $j,k=1,2$ and $j \neq k$, in Equation (\ref{TransRates}) are chosen such that they represent a clear seasonal pattern, with higher intensities to move from state 2 to state 1 in spring, and vice versa in autumn (cf.\ Figure~\ref{fig_sim} in Appendix B).
The death rates are the same for both states and constant over time, with an expected survival time of about 22 years.

When simulating the data, we use one value of $y(t)$ for each day in Equation (\ref{TransRates}), leading to transition intensities that are constant over a day.
For the parameter estimation, we then approximate the continuously varying $y(t)$ in Equation (\ref{TransRates}) by the midpoint $c_r$ of the interval $\tau_r$ within which $t$ lies (cf.\ Equation (\ref{piecewiseTransRates}) in Section \ref{s_apply}). 
We first repeatedly estimate the model parameters for a \textit{single} data set of capture histories, numerically maximising the likelihood given in Equation (\ref{llk_cov}), only varying the length $l$ of the intervals used for the likelihood approximation, considering the Fibonacci sequence $l = 2,3,5,8,13,21,34,55,89$.
The computation times as well as the maximum log-likelihood values obtained with the different interval lengths are shown in Table~\ref{tab_sim}.
The computation time increases with decreasing interval length, whereas the maximum likelihood value is roughly the same for all $l \leq 21$, and virtually identical for all $l \leq 5$. 
This is to be expected:\ given a sufficiently fine piecewise constant approximation, a further decrease in the interval lengths does not yield a relevant difference in the likelihood calculation (``diminishing returns''). 
For this particular simulation setting, choosing $8 \leq l \leq 21$ seems to provide a good balance between approximation accuracy and computational cost.
\begin{table}[!b]
    \centering
    \caption{Computation times and maximum log-likelihood (abbr.\ llk) values in the simulation experiment for different interval lengths $l$ used for the likelihood approximation.}
    \label{tab_sim}
    \begin{tabular}{ccc}
         Int.\ length $l$ & Comp.\ time (min) & -- llk \\ \hline
         89 & 38.91 & 32341.01 \\
         55 & 41.00 & 32335.63 \\
         34 & 43.78 & 32333.71 \\
         21 & 46.75 & 32333.21 \\
         13 & 50.15 & 32333.17 \\
         8 & 51.02 & 32333.36 \\
         5 & 61.19 & 32333.24 \\
         3 & 73.86 & 32333.25 \\
         2 & 95.52 & 32333.24 \\
    \end{tabular}
\end{table}
The parameter estimates obtained using interval lengths of $l>21$ in the approximation do in fact hardly differ from those obtained using a smaller $l$ (cf.\ Table~\ref{tab_simSI} in Appendix B), meaning that even a relatively coarse partition of the observed time interval $[0,t_T]$ leads to satisfactory results.

In a second simulation study, we investigated the effect of the sample size $n$ on the estimation results. 
The setting is the same as before, only that we reduced the length of the capture histories to three years.
For each of the three sample sizes $n=100, 200, 400$, we simulated 100 data sets and estimated the model parameters fixing the interval length used in the approximation at $l=20$. 
\begin{figure}[!b]
    \centering
    \includegraphics[width=137mm]{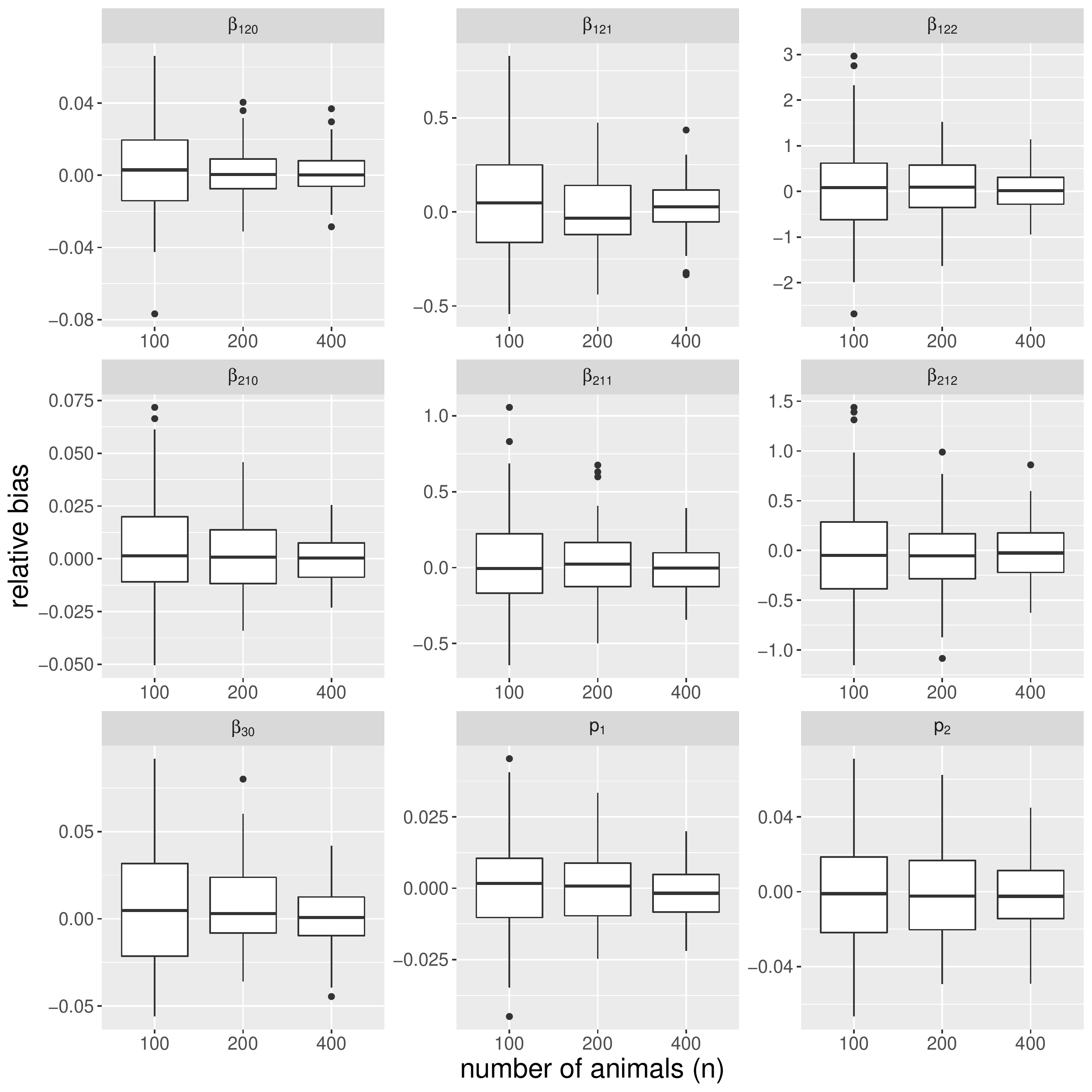}
    \caption{Boxplots of relative bias of the estimated model parameters from 100 simulation runs for the number of animals $n=100, 200, 400$.}
    \label{fig_bias}
\end{figure}
From the 100 simulation runs, we then calculated the relative bias $\text{RB}=\frac{\hat{\theta}-\theta}{\theta}$ for each model parameter; the results are displayed in Figure~\ref{fig_bias}. 
The parameter estimators appear to be approximately unbiased already for $n=100$, with the variance decreasing as $n$ increases. 
Overall, the true parameters could be recovered fairly well in this setting.

\section{Discussion}
\label{s_discuss}

When modelling the evolution of systems over time, a high-level conceptual decision is whether to use a discrete-time or a continuous-time model formulation.
While discrete-time models tend to be easier to work with, some data sets may in fact require a continuous-time model formulation. 
Moreover, as time series models generally need to be interpreted with respect to the sampling rate, there may also be conceptual advantages of a continuous-time approach.
Corresponding continuous-time models have been developed in different areas including medicine (e.g.\ Jackson et al., 2003; Williams et al., 2019), finance (e.g.\ Barndorff-Nielsen and Shephard, 2001; Krishnamurthy, Leoff, and Sass, 2018), and biology (e.g.\ Minin and Suchard, 2008; Fuchs, 2013).

Driven by advances in data collection, continuous-time models have recently become increasingly popular within ecology.
Specifically in movement ecology, continuous-time models are appealing as they can accommodate irregular time intervals between observations --- as often encountered for example when tracking marine mammals (see, e.g., Jonsen, Flemming, and Myers, 2005; van Beest et al., 2019) --- and are generally independent of the temporal resolution at which observations are made (McClintock et al., 2014; Patterson et al., 2017; Michelot and Blackwell, 2019).
In the context of distance sampling, relaxing the commonly made assumption of animals being stationary during a survey can be accomplished by modelling their movement or availability process in continuous time.
For example, Langrock et al.\ (2013) and Borchers and Langrock (2015) account for marine mammals' availability by modelling their surfacing events --- occurring irregularly and clustered in time --- as a Markov-modulated Poisson process, while Glennie et al.\ (2017) account for animal movement within a distance sampling survey by incorporating a continuous space-time model for the animal's unobserved trajectory.
In spatial capture-recapture studies, surveys commonly operate in continuous time and provide exact detection times as obtained for example using microphones or motion-activated cameras.
In order to use the full information contained in such data, continuous-time methods are necessary (see for example Borchers et al., 2014).

In contrast to the ecological scenarios described above, capture-recapture studies are usually designed according to regular sampling schemes, i.e.\ with regularly spaced capture occasions (e.g.\ once per year).
However, for practical considerations it is sometimes not possible to sample regularly, for example due to resource issues or environmental conditions (as for the bottlenose dolphin data considered in this work).
In such instances, fitting the traditional discrete-time AS model often leads to additional model complexity to account for such irregularity.
For example, Dupuis (1995) and King and Brooks (2002) consider data relating to lizards (\textit{Lacerta vivipara}) where the capture occasions correspond to three years of data, with the data collected in June and August. 
Such survey designs necessarily lead to the need for time-dependent AS model parameters to account for the time differentials, thus generally increasing the model complexity, decreasing the biological interpretability, and decreasing parameter precision. 
Further, capture occasions may span a period of time (from days to months), as for example in King and Brooks (2003) where the considered data are recorded annually, with the data collection occurring from June to August. 
For the discrete-time AS model to be applicable in such cases, the data may have to be reduced to a single observation within the whole sampling period, instead of using the finer-level information.
The continuous-time AS model proposed within this manuscript naturally addresses these issues, without the need to add potential additional complexity. 
It thus constitutes the natural modelling framework for dealing with irregularly sampled multi-state capture-recapture data.

\section*{Acknowledgements}

We thank Paul Thompson and Barbara Cheney from the University of Aberdeen Lighthouse Field Station and Phil Hammond and Monica Arso-Civil for provision of the dolphin data used in this study. We also thank all organisations who have provided funding over the three decades of research that has contributed to the individual based bottlenose dolphin project. All photo-identification surveys were conducted under Scottish Natural Heritage Animal Scientific Licences.
We also thank Andrea Langrock for designing the illustration of the dolphin's capture history in Figure~\ref{fig_capHist}.
Finally, we thank Julia Schemm and Irina Janzen for their contributions to a preliminary analysis of the dolphin data which motivated the present paper.

\def\bibindent{24pt}

\newpage

\section*{Appendix}

\subsection*{Appendix A: Additional information on the results of movement patterns for the motivating data}

\begin{table}[h]
    \centering
    \caption{Estimation results and 95\% confidence intervals for the motivating capture-recapture data of bottlenose dolphins. The model parameters are the transition intensities $\beta_{jk0}^{(z)}$, $\beta_{jk1}^{(z)}$, and $\beta_{jk2}^{(z)}$ for $j,k=1,2$ and $j \neq k$ (with $z=0$ for females and $z=1$ for males), the intercept $\beta_{30}$ and difference $\beta_{31}$ in mortality rates and the detection probabilities $p_1$ and $p_2$ for SAC and T\&F, respectively. }
    \label{tab_resSI}
    \begin{footnotesize}
        \begin{tabular}{l c c c c c c}
         & $\beta_{120}^{(0)}$ & $\beta_{121}^{(0)}$ & $\beta_{122}^{(0)}$ & $\beta_{210}^{(0)}$ & $\beta_{211}^{(0)}$ & $\beta_{212}^{(0)}$  \\ \hline
         estimate & -6.855 & -0.816 & -0.752 & -7.529 & -0.229 & -2.274 \\ \vspace{3mm}
         95\% CI & [-7.09; -6.62] & [-1.14; -0.49] & [-1.03; -0.48] & [-7.89; -7.16] & [-0.52; 0.06] & [-2.74; -1.81] \\ 
         & $\beta_{120}^{(1)}$ & $\beta_{121}^{(1)}$ & $\beta_{122}^{(1)}$ & $\beta_{210}^{(1)}$ & $\beta_{211}^{(1)}$ & $\beta_{212}^{(1)}$ \\ \hline
         estimate & -6.886 & -1.293 & -0.942 & -7.413 & -0.191 & -2.490 \\ \vspace{3mm}
         95\% CI & [-7.22; -6.55] & [-1.68; -0.90] & [-1.27; -0.62] & [-7.82; -7.01] & [-0.48; 0.01] & [-2.99; -1.99] \\
         
         & $p_1$ & $p_2$ & $\beta_{30}$ & $\beta_{31}$ &  & \\ \hline
         estimate & 0.201 & 0.191 & -9.403 & 0.084 &  & \\ 
         95\% CI & [0.196; 0.205] & [0.183; 0.199] & [-9.70; -9.10] & [-0.35; 0.52] &  & 
        \end{tabular}
    \end{footnotesize}
\end{table}

\newpage

\subsection*{Appendix B: Additional information for the simulation experiments}

\begin{figure}[!htb]
    \centering
    \includegraphics[width=120mm]{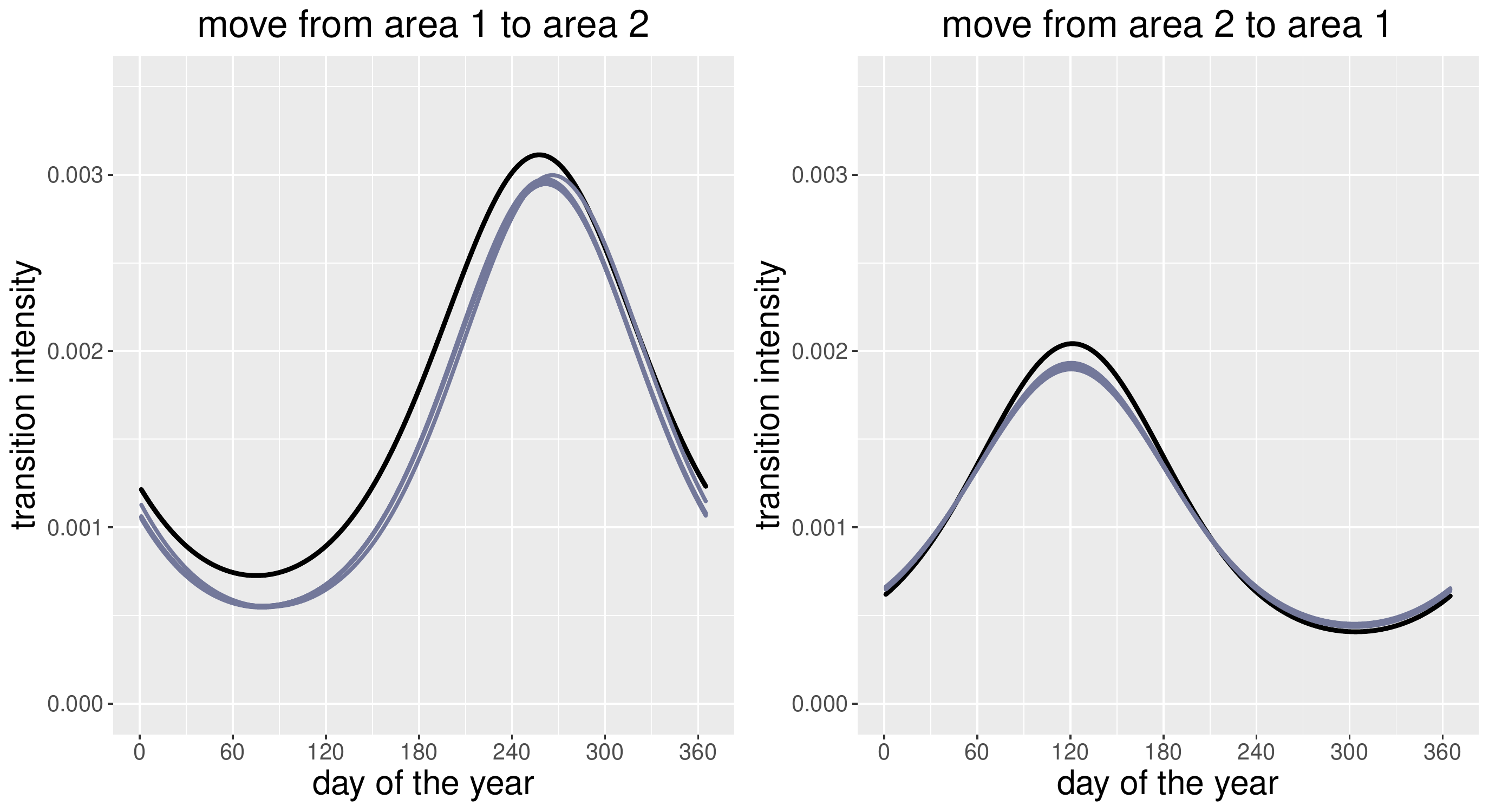}
    \caption{True (black) and estimated (coloured lines) transition intensities of the simulation experiment as a function of the covariate \textit{day of the year}. The left plot shows the intensities to move from area 1 to area 2 and the right plot vice versa. The estimated intensities for different interval lengths used for the likelihood approximation are almost identical visually. Coefficients underlying this figure are provided in Table~\ref{tab_simSI}.}
    \label{fig_sim}
\end{figure}

\begin{table}[!htb]
    \centering
    \caption{True and estimated parameter values in the simulation experiment for different interval lengths $l$ used for the likelihood approximation. For $l \leq 8$ the estimated parameter values do not change anymore (up to third decimal). }
    \label{tab_simSI}
    \begin{footnotesize}
        \begin{tabular}{cccccccccc}
         Int. length & $\beta_{120}$ & $\beta_{121}$ & $\beta_{122}$ & $\beta_{210}$ & $\beta_{211}$ & $\beta_{212}$ & $\beta_{30}$ & $p_1$ & $p_2$ \\ \hline
         89 & -6.658 & -0.841 & -0.113 & -6.984 & 0.627 & -0.346 & -9.001 & 0.401 & 0.198 \\
         55 & -6.666 & -0.830 & -0.179 & -6.992 & 0.647 & -0.359 & -9.001 & 0.401 & 0.198 \\
         34 & -6.662 & -0.819 & -0.169 & -6.994 & 0.653 & -0.361 & -9.001 & 0.401 & 0.198 \\
         21 & -6.664 & -0.821 & -0.176 & -6.989 & 0.640 & -0.354 & -9.001 & 0.401 & 0.198 \\
         13 & -6.663 & -0.821 & -0.171 & -6.988 & 0.638 & -0.351 & -9.001 & 0.401 & 0.198 \\
         8 & -6.662 & -0.818 & -0.172 & -6.988 & 0.637 & -0.350 & -9.001 & 0.401 & 0.198 \\
         5 & -6.662 & -0.818 & -0.172 & -6.988 & 0.637 & -0.350 & -9.001 & 0.401 & 0.198 \\
         3 & -6.662 & -0.818 & -0.172 & -6.988 & 0.637 & -0.350 & -9.001 & 0.401 & 0.198 \\
         2 & -6.662 & -0.818 & -0.172 & -6.988 & 0.637 & -0.350 & -9.001 & 0.401 & 0.198 \\
         true values & -6.5 & -0.7 & -0.2 & -7.0 & 0.7 & -0.4 & -9 & 0.4 & 0.2
    \end{tabular}
    \end{footnotesize}
\end{table}

\end{spacing}

\end{document}